\documentclass[aps,pra,twocolumn,superscriptaddress,showpacs,longbibliography]{revtex4-1}

\usepackage{graphicx}
\usepackage{amsmath}
\usepackage{color}

\usepackage[bookmarks=true,colorlinks=true,urlcolor=blue,linkcolor=blue,citecolor=blue,breaklinks]{hyperref}

\usepackage{cleveref}
\usepackage{amssymb}
\usepackage{booktabs}

\begin{document}


\title{Phonon-assisted processes in the ultraviolet transient optical response of graphene}





\newcommand*{\DIPC}[0]{{
Donostia International Physics Center (DIPC),
Paseo Manuel de Lardizabal 4, 20018 Donostia-San Sebasti\'an, Spain}}

\newcommand*{\IFS}[0]{{
Center of Excellence for Advanced Materials and Sensing Devices, Institute of Physics, Bijeni\v{c}ka 46,
10000 Zagreb, Croatia}}

\author{Dino Novko}
\email{dino.novko@gmail.com}
\affiliation{\IFS}
\affiliation{\DIPC}

\author{Marko Kralj}
\affiliation{\IFS}

\begin{abstract}

Many recent experiments investigated potential and attractive means of modifying many-body interactions in two-dimensional materials through time-resolved spectroscopy techniques. However, the role of ultrafast phonon-assisted processes in two-dimensional systems is rarely discussed in depth. Here, we investigate the role of electron-phonon interaction in the transient optical absorption of graphene by means of first-principles methods. It is shown at equilibrium that the phonon-assisted transitions renormalize significantly the electronic structure. As a result, absorption peak around the Van Hove singularity broadens and redshifts by around 100\,meV. In addition, temperature increase and chemical doping are shown to notably enhance these phonon-assisted features. In the photoinduced transient response we obtain spectral changes in close agreement with the experiments, and we associate them to the strong renormalization of occupied and unoccupied $\pi$ bands, which predominantly comes from the coupling with the zone-center $E_{2g}$ optical phonon. Our estimation of the Coulomb interaction effects shows that the phonon-assisted processes can have a dominant role even in the subpicosecond regime.

\end{abstract}

\maketitle



\section*{Introduction}

The ability of altering the electronic structure as well as corresponding excitation spectrum is one of the most impressive feature of two-dimensional and layered materials\,\cite{bib:castroneto09,bib:wang12}. Electronic properties of such systems are shown to be highly sensitive to external perturbation that introduces excess electron or hole concentrations. For instance, electrostatic and chemical doping techniques were successfully utilized in order to tune optical absorption as well as plasmon and exciton energies in graphene-based materials\,\cite{bib:ju11,bib:fei12,bib:zhou12,bib:khrapach12,bib:bao14,bib:mak14,bib:hage18}, transitions metal dichalcogenides\,\cite{bib:mouri13,bib:wang13,bib:tongay13,bib:steinhoff14,bib:liang15,bib:chernikov15a,bib:yao17}, and corresponding van-der-Waals heterostructures\,\cite{bib:ugeda14,bib:woessner15}.

In addition to these static modifications, it is of utter importance to comprehend transient electronic structure changes in time domain. Many pump-probe spectroscopy techniques (e.g., ultrafast photoemission and optical absorption) were performed on a variety of two-dimensional materials in order to gain such an insight\,\cite{bib:winnerl11,bib:obraztsov11,bib:pagliara11,bib:roberts14,bib:chernikov15,bib:pogna16,bib:ulstrup16,bib:ruppert17,bib:sie17,bib:cunningham17,bib:chi18,bib:miao18}. In these time-resolved experiments, it is believed that subpicosecond regime is dominated by many-body Coulomb interactions between photoinduced electrons, which in turn renormalizes the electronic band structure\,\cite{bib:pogna16}. The latter process is then followed by the electron-phonon scattering events, leading the material to equilibrium state\,\cite{bib:ruppert17}. In fact, the sequence of fast and slow relaxation events at two characteristic time scales of transient response associated with electron and phonon-assisted dynamics, respectively (the so-called biexponential decay), appears to be universal and was observed in various photoemission and optical absorption experiments (e.g., in photemission intensity of MoS$_2$ valence bands\,\cite{bib:ulstrup16} and in transient reflectivity spectra of graphite\,\cite{bib:pagliara11}).

Despite these remarkable experimental studies, our understanding of transient response of two-dimensional materials is still incomplete. First of all, most of the studies are focused only on the aspect of the Coulomb interactions and ensuing subpicosecond electronic structure changes\,\cite{bib:pogna16,bib:ulstrup16,bib:cunningham17}. A robust many-body theoretical framework describing these interactions (i.e., electron-electron and electron-hole) is quite well established for both equilibrium\,\cite{bib:yang11,bib:mak14,bib:steinhoff14} and nonequilibrium conditions\,\cite{bib:perfetto15,bib:pogna16,bib:molinasanchez17}. For example, such a theory had successfully explained the electronic band gap renormalization of MoS$_2$, observed in transient absorption spectroscopy, in terms of Coulomb interaction between photoexcited electron-hole pairs\,\cite{bib:pogna16}. On the other hand, the impact of the phonon dynamics on the transient optical response and the related band gap renormalization is rarely discussed in depth, even though the experimental results display its relevance (aforesaid slow decay process)\,\cite{bib:pagliara11,bib:roberts14,bib:ulstrup16,bib:ruppert17,bib:miao18}. The intensity of the transient response associated with this slow decay mechanism is believed to be smaller than the subpicosecond response ruled by electron interactions. Nevertheless, it is still considered significant and it lasts for longer time\,\cite{bib:roberts14}, which could be exploited whenever the stable and long-lived electronic structure changes are of practical usage. All in all, gaining a full microscopic insight, i.e., both electronic and lattice aspects, of photoinduced transient response is of pivotal importance for optimization and control of optoelectronic devices based on two dimensional materials.

Regarding the role of phonons in transient optical absorption, the high-energy (UV) response of graphene stands as a quite interesting, but unresolved issue\,\cite{bib:pagliara11,bib:mak14,bib:roberts14,bib:binder16,bib:yang17}. Recently, the time-resolved differential optical transmission study on graphene had shown a considerable band structure renormalization around the Van Hove singularity (VHs) point on the picosecond time scale (i.e., $\tau>10$\,ps)\,\cite{bib:roberts14}. This enduring transient response was ascribed to electron-phonon coupling (EPC), where acoustic modes were believed to play a dominant role. Interestingly, along with the subpicosecond transient response ruled by the many-body Coulomb interactions, a similar long-lived decay process around the VHs was observed as well in pump-probe optical absorption spectroscopy of graphite\,\cite{bib:pagliara11}. However, when it comes to the theoretical considerations, most of them are dealing with the low-energy (THz) transient optical response\,\cite{bib:butscher07,bib:malard13,bib:kadi14,bib:tomadin18} or they are parameter-based\,\cite{bib:roberts14,bib:binder16}, while the only available \emph{ab initio} studies of phonon-assisted optical absorption in high-energy regime of graphene are for equilibrium condition\,\cite{bib:mak14,bib:yang17}. Even more, recent study of phonon impact on optical absorption is conflicting with experiments, showing only an increase of absorption around the VHs and no band renormalization due to EPC\,\cite{bib:yang17}. All these calls for a thorough and quantitative study on the transient optical response of graphene that could resolve this issue and unveil the corresponding role of phonon-assisted processes.

Here we utilize density functional and density functional perturbation theories in order to investigate the impact of EPC on optical absorption of graphene. In particular, the study combines the theoretical framework of the electromagnetic linear response\,\cite{bib:novko16} with quasiclassical Williams-Lax approach for incorporating temperature-dependent phonon-induced band renormalization processes in optical absorption\,\cite{bib:zacharias15,bib:zacharias16,bib:giustino17}. Additionally, photoinduced transient optical response is simulated by introducing transient phonon temperatures, as obtained from the effective temperature model\,\cite{bib:allen87,bib:lui10,bib:johannsen13} based in first principles\,\cite{bib:caruso19}, into the temperature-dependent optical absorption formula.

The results on the equilibrium optical absorption show that the inclusion of the EPC renormalizes the band structure and thus redshifts (by around 100\,meV) and broadens the high-energy absorption peak arising from the vertical transitions between occupied and unoccupied $\pi$ bands around VHs (i.e., at the M point of the Brillouin zone). Consequently, the optical absorption just below and above the VHs is increased and decreased, respectively, with regards to the bare spectrum. Semiquantitative analysis shows that the band structure renormalization is mostly due to coupling between electrons and longitudinal optical phonon modes at the Brillouin zone center, while the acoustic modes only play a minor role. This is contrary to the conclusions extracted from Ref.\,\cite{bib:roberts14}, where it was argued that the coupling between electrons and acoustic phonons is in fact the main cause of band gap renormalization in graphene. Furthermore, the increase of the temperature enhances the effect of the EPC and in turn redshifts the VHs peak and decreases its intensity even more. The impact of the excess electron concentration on the temperature-dependent optical absorption is also investigated by adsorbing lithium atoms or replacing carbon atoms with nitrogen. This modification further decreases the intensity and redshifts the VHs peak, which is in line with the recent electron energy loss spectroscopy experiment on nitrogen-doped graphene\,\cite{bib:hage18}. 

Our simulations of the photoinduced transient response show a time-dependent change of absorption in the vicinity of the VHs-peak that is in a very good agreement with the experiment\,\cite{bib:roberts14}. We also analyze how the transient optical absorption and band structure of graphene are altered with the change of the pump laser intensity and with the change of the electron doping concentration. Since electron correlations have a significant impact on the VHs-peak of graphene\,\cite{bib:yang09,bib:kravets10,bib:yang11,bib:mak11,bib:chae11,bib:malic11,bib:mak14}, we also deliver a qualitative estimation on the role of the Coulomb interaction and excitonic effects in transient optical absorption of graphene.

The role of the EPC in transient response of graphene is disentangled and shown to be crucial for understanding the pathways of manipulating the optical response and band structure renormalization in graphene-based materials.

\section*{Results}

\textbf{Phonon-assisted absorption.--} Quite standard procedure for simulating the optical absorption spectrum of some material is to calculate the corresponding dynamical conductivity, in particular, its real part in the long-wavelength limit, which we denote here as $\sigma(\omega)$. For noninteracting electrons, i.e., when the EPC is omitted, as it is the case in the density functional theory, $\sigma(\omega)$ includes only momentum- and energy-conserving (vertical) electronic transitions. In other words, $\sigma(\omega)$ only reflects the ground state electron density of states and the energy of the interband gap is temperature independent. On the other hand, when the EPC is included, the optically excited electron can scatter on the phonons, whereby its mementum and energy are changed. This results in the phonon-assisted absorption and the accompanying temperature-dependent band structure renormalization.

The incorporation of the EPC into optical absorption formula $\sigma(\omega)$ is very often done by means of the many body perturbation theory, where electron-hole pairs are dressed with the EPC\,\cite{bib:giustino17}. The latter can be realized by correcting the electron energies in $\sigma(\omega)$ with single-electron self-energy due to EPC\,\cite{bib:allen76}, by expanding the Fermi's golden rule expression of $\sigma(\omega)$ up to second order\,\cite{bib:noffsinger12}, or even by solving the Bethe-Salpeter-type equation for the EPC\,\cite{bib:chakraborty78,bib:antonius17}. Instead of using these robust but numerically demanding theories, in this work we calculate the phonon-assisted absorption with the quasiclassical Williams-Lax approach where noninteracting $\sigma(\omega)$ is statistically averaged over the structures distorted along the eigenvectors of the system phonon modes\,\cite{bib:zacharias15,bib:zacharias16,bib:giustino17}. It was in fact shown that this method requires only a single distorted structure per given temperature provided that the corresponding supercell is large enough to account for all the relevant phonon modes\,\cite{bib:zacharias16}. The optical absorption formula for the distorted structure is then calculated as\,\cite{bib:novko16}
\begin{eqnarray}
\sigma^{\tau}(\omega)=\frac{2}{\Omega}\mathrm{Re}\sum_{\mathbf{k},nm}\frac{i\left|J^{\tau}_{nm\mathbf{k}}\right|^2}{\varepsilon^{\tau}_{m\mathbf{k}}-\varepsilon^{\tau}_{n\mathbf{k}}}\,\frac{f^{\tau}_{n\mathbf{k}}-f^{\tau}_{m\mathbf{k}}}{\omega+\varepsilon^{\tau}_{n\mathbf{k}}-\varepsilon^{\tau}_{m\mathbf{k}}+i\eta},
\label{bib:eq1}
\end{eqnarray}
where $n$ and $m$ are the band indices, $\mathbf{k}$ is the electron momentum, $\varepsilon^{\tau}_{n\mathbf{k}}$ is the electron energy, $J^{\tau}_{nm\mathbf{k}}$ is the current vertex function, $f^{\tau}_{n\mathbf{k}}$ is the Fermi-Dirac distribution function, and $\Omega$ is the unit cell volume. The index $\tau$ indicates which quantities are altered by the temperature-dependent displacement of the atoms in the supercell along $\Delta\tau$ (see Methods section for further details).

\begin{figure}[t]
\includegraphics[width=0.46\textwidth]{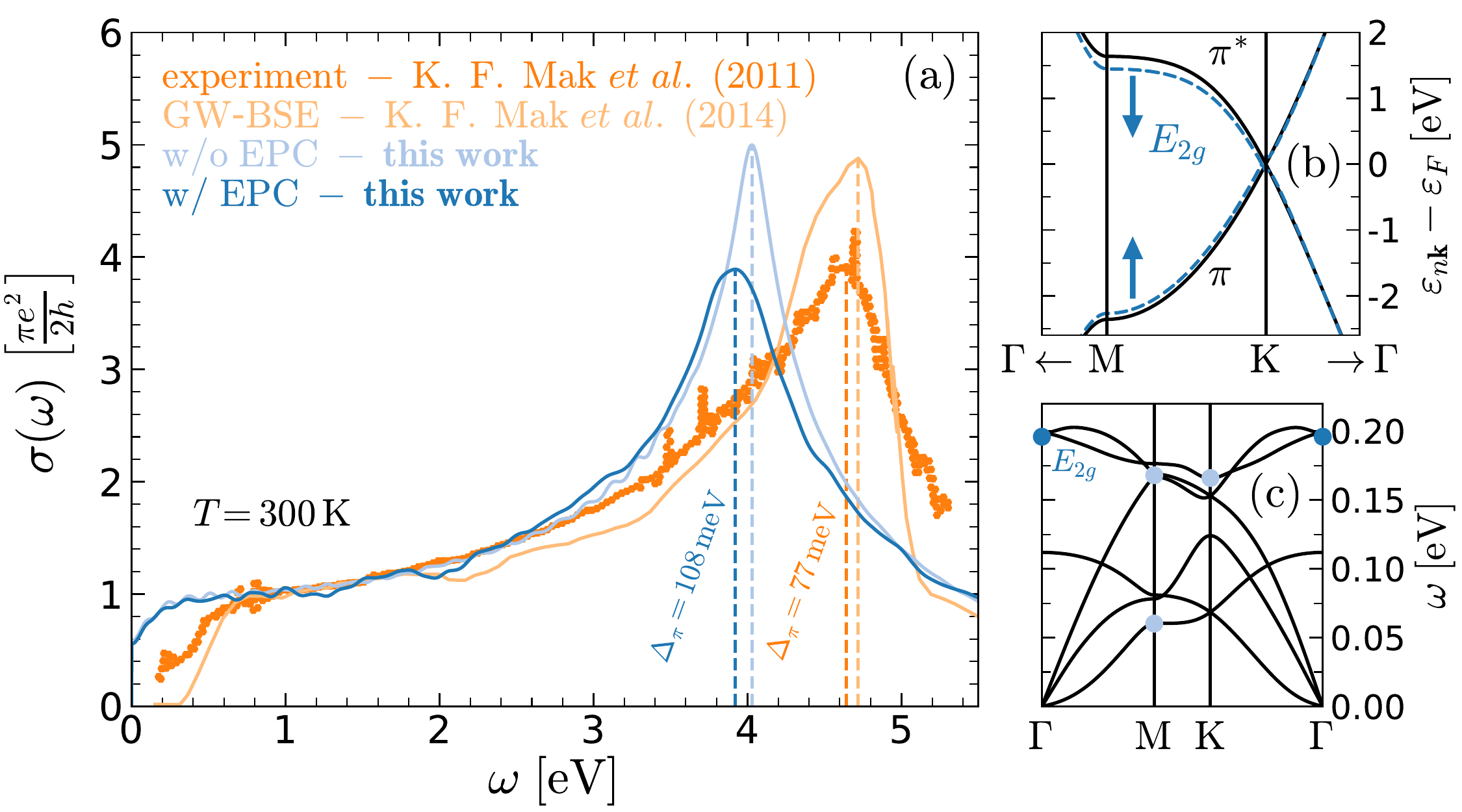}
\caption{\label{fig:fig1}Impact of electron-phonon coupling on optical response of graphene. (a) Dark and light blue lines show the optical absorption $\sigma(\omega)$ of graphene in equilibrium state ($T=300$\,K) with and without EPC effects included, respectively. Dark orange dots are the experimental measurements from Ref.\,\cite{bib:mak11} and light orange line is the theoretical result obtained in Ref.\,\cite{bib:mak14}, which includes both electron-electron (GW) and electron-hole (Bethe-Salpeter equation, BSE) interactions. (b) Electron band structure of graphene distorted along the eigenvectors of $E_{2g}$ phonon mode (blue dashed lines) compared with undistorted graphene (black lines). (c) Phonon band structure of graphene. Dark blue dots highlight the $E_{2g}$ mode, while light blue dots point the LO and ZA modes at M point, and the TO mode at K point.}
\end{figure}

In Fig.\,\ref{fig:fig1}(a) we show the results of Eq.\,\eqref{bib:eq1} calculated for the pristine graphene in the equilibrium state (i.e., both electrons and phonons have $T=300$\,K) (dark blue) along with the results of the noninteracting $\sigma(\omega)$ (light blue). First of all, we note that the optical absorption of pristine graphene is characterized by almost constant value at low energies and with the distinctive peak at $\varepsilon_{\pi}\approx4$\,eV, originating from the transitions  between  occupied  and  unoccupied $\pi$ bands  around  the VHs point\,\cite{bib:novko15,bib:kupcic14} [see Fig.\,\ref{fig:fig1}(b)]. The inclusion of the EPC in the optical absorption clearly modifies the energy window around $\varepsilon_{\pi}$. In particular, the VHs peak position redshifts by $\Delta_{\pi}=108$\,meV, while the intensity decreases by $\Delta I_{\pi}\approx\pi e^2/2h$. These changes are actually in line with the general conclusions on the impact of the EPC on the band structure\,\cite{bib:allen76,bib:chakraborty78,bib:giustino17}. However, these results contradict the recent first-principles study of the phonon-assisted absorption in graphene where seemingly similar quasiclassical approach was used\,\cite{bib:yang17}. Their results show an overall increase of the optical absoprtion around the VHs peak due to EPC without energy renormalization. Another study considered the EPC effects in the optical absorption of graphene by introducing the broadening of the electronic states due to EPC\,\cite{bib:mak14}. This produced a decrease and widening of the VHs peak, in good agreement with our result. Our result of $\sigma(\omega)$ with EPC concurs quite well also with the experiment\,\cite{bib:mak11}, if the energy shift of about 0.4\,eV due to combination of electron-electron and electron-hole interactions is omitted [see Fig.\,\ref{fig:fig1}(a)].
This blueshift due to carrier-carrier interactions is nicely depicted in Fig.\,3(a) of Ref.\,\cite{bib:yang11}.
In fact, by comparing the optical absorption obtained through Bethe-Salpeter equation (electron-hole interaction) with GW-corrected energies (electron-electron interaction), but without the EPC\,\cite{bib:mak14}, and the experimental spectrum\,\cite{bib:mak11}, the resulting discrepancies are in line with the changes that we get by including the EPC, i.e., $\Delta_{\pi}=77$\,meV and $\Delta I_{\pi}\approx\pi e^2/2h$. The latter gives us confidence that Eq.\,\eqref{bib:eq1} accounts for the EPC effects in the graphene optical spectrum in a proper way.

\begin{figure}[b]
\includegraphics[width=0.48\textwidth]{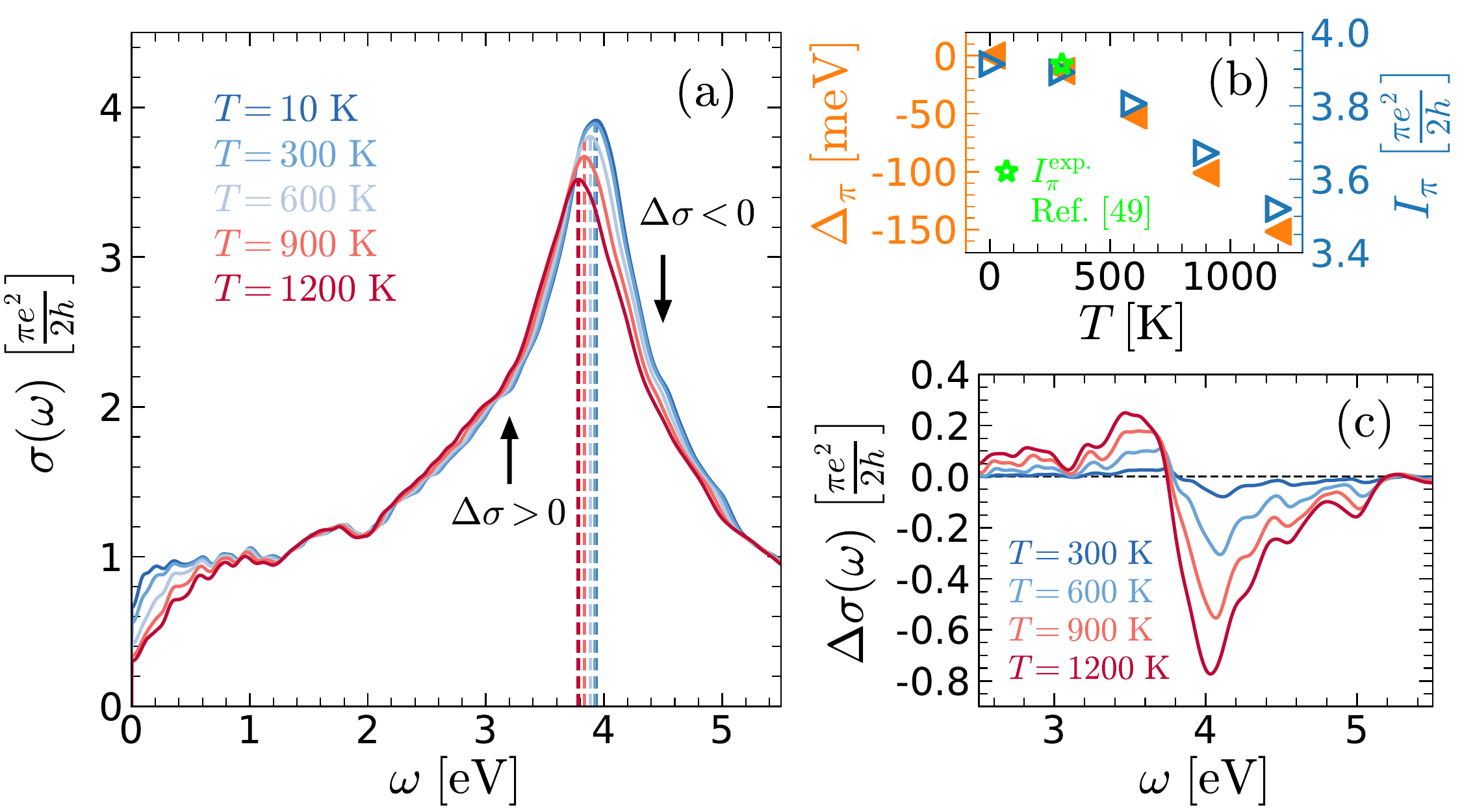}
\caption{\label{fig:fig2}Temperature effects in absorption spectra of graphene. (a) Optical absorption $\sigma(\omega)$ of graphene with EPC effects included in equilibrium condition for different temperatures in the range of $T=10-1200$\,K. (b) Energy shifts of the VHs peak $\Delta_{\pi}$ (orange triangles) and the corresponding intensity $I_{\pi}$ (blue triangles) as a function of temperature. The experimental intensity $I_{\pi}^{\mathrm{exp.}}$ at 300\,K is also shown (green star)\,\cite{bib:mak11}. (c) Differential absorption $\Delta\sigma(\omega)$ for several temperatures. The reference optical absorption is taken at $T=10$\,K.}
\end{figure}

\begin{figure*}[t]
\centering
\includegraphics[width=0.9\textwidth]{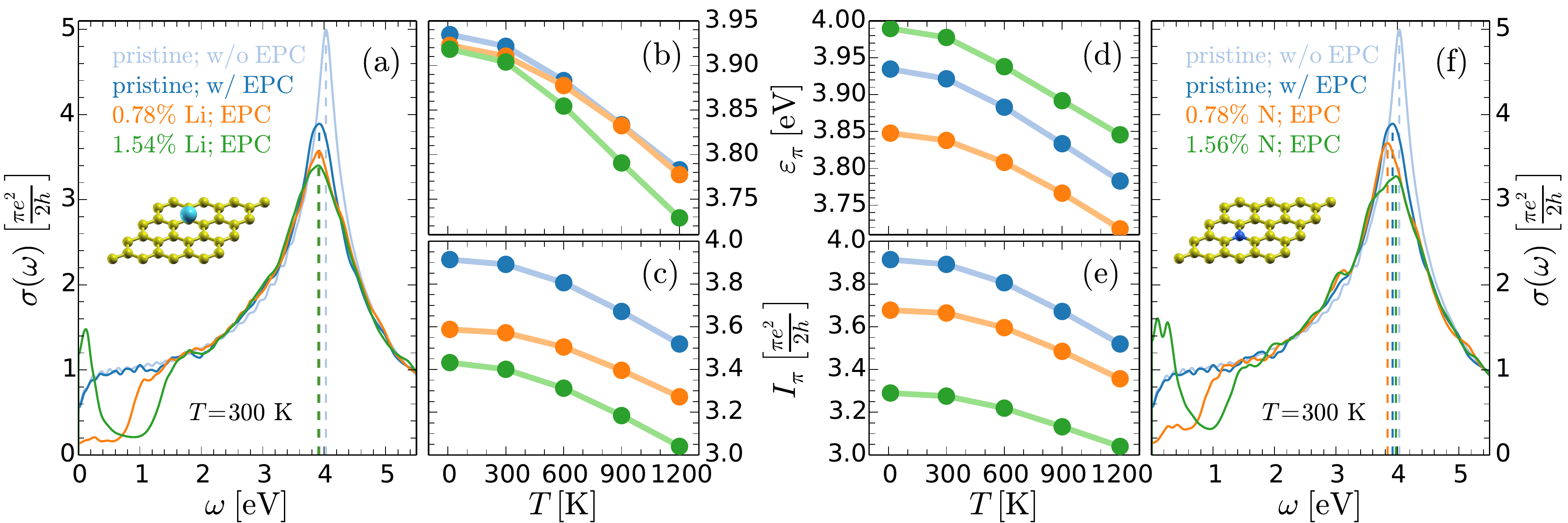}
\caption{\label{fig:fig3}Role of electron-phonon coupling in optical response of doped graphene. (a) Optical absorption $\sigma(\omega)$ for pristine (dark blue) and doped graphene with 0.78\% (orange) and 1.54\% (green) Li concentrations with EPC effects included ($T=300$\,K). Optical absorption for pristine graphene without EPC effects is shown in light blue. (b) Energy $\varepsilon_{\pi}$ and (c) intensity $I_{\pi}$ of the VHs peak as a function of temperature is shown for pristine (blue circles) and doped graphene with 0.78\% (orange circles) and 1.54\% (green circles) Li concentrations. (d),(e),(f) Same as in (a),(b),(c), but for the N-doped graphene.}
\end{figure*}

We proceed with the qualitative mode-resolved analysis of the EPC band structure renormalization. By distorting the supercell structure along the eigenvector of the single phonon mode we can dissect which modes in particular are responsible for the redshift of the interband gap in the graphene electronic band structure [see Fig.\,\ref{fig:fig1}(b)]. In Fig.\,\ref{fig:fig1}(c) we show the full phonon band structure and highlight the specific modes responsible for the band shifts. The dark blue dots underline the LO mode at $\Gamma$ point of the Brillouin zone (the $E_{2g}$ mode), which is most responsible for the shift. The light blue dots show the modes that give a minor contribution to the shift. These include the LO mode at M point, the TO mode at K point ($A_1'$), and the ZA mode at M point. The contributions of the rest of the modes are insignificant. These results are in agreement with the theoretical calculations of the EPC strength and matrix elements\,\cite{bib:piscanec04}, where it was shown that the $E_{2g}$ and $A_1'$ modes are much more strongly coupled to the electrons in graphene than the rest of the modes. This anisotropy of the EPC in graphene\,\cite{bib:kampfrath05,bib:malic11,bib:chae11,bib:berciaud10,bib:caruso19} comes primarily from the reduced phase space of Dirac-cone-type band structure, where mostly intravalley ($E_{2g}$-electron) and intervalley ($A_1'$-electron) scatterings are preferred. On the contrary, in the time-resolved differential optical transmission experiment the transient band shifts in graphene were mostly attributed to the acoustic phonons, while the relevance of the optical phonons was disregarded due to the estimated low phonon temperature, i.e., lower than the phonon energy\,\cite{bib:roberts14}. Below, we will show that the $E_{2g}$ mode still underlies the EPC-induced band shifts in the transient condition and that the phonon temperatures can be as high as 1200\,K.

\textbf{Temperature effects and electron doping.--} As the temperature increases the displacements of the atoms in the distorted graphene are larger and thus the EPC-induced changes in the optical absorption become more pronounced. In Fig.\,\ref{fig:fig2}(a) we display the optical absorption spectra of graphene for different temperatures in the range of $T=10-1200$\,K. With the elevated temperature both the energy $\varepsilon_{\pi}$ and the intensity $I_{\pi}$ of the VHs decreases. In the presented temperature range, $\varepsilon_{\pi}$ decreases by around 150\,meV and $I_{\pi}$ is reduced by $0.5\pi e^2/2h$ [see Fig.\,\ref{fig:fig2}(b)]. Such a reduction of the interband gap with increasing temperature goes in line with the theory of the EPC-induced band gap renormalization in semiconductors\,\cite{bib:allen76,bib:giustino17}. As a result of the reduction of $\varepsilon_{\pi}$, the differential absorption $\Delta\sigma(\omega)$ [with the reference absorption $\sigma(\omega)$ at $T=10$\,K] increases for $\omega\lesssim \varepsilon_{\pi}$ and decreases for $\omega\gtrsim\varepsilon_{\pi}$ [see Fig.\,\ref{fig:fig2}(c)]. Very similar spectral changes were obtained for graphene\,\cite{bib:roberts14} and graphite\,\cite{bib:pagliara11} by using time-resolved pump-probe spectroscopy, however, we leave the discussion on the transient optical response for later. Note that the decrease of the absorption for $\omega<1$\,eV is due to the increase of the occupied states in the upper $\pi$ band with elevated temperature, which in turn blocks the interband electronic transitions, and has nothing to do with the EPC effects.

The effects of the EPC can also be altered with the electron doping techniques. For example, it was shown in the photoemission spectroscopy study that the EPC strength enhances with the increase of the excess electron concentration provided by the dopant atom\,\cite{bib:fedorov14}, which was also confirmed with the first-principles calculations\,\cite{bib:novko17}. The dopant atom that increases the electron concentration in graphene can be intercalated between the substrate and graphene sheet (usually alkali and alkaline earth metal atoms)\,\cite{bib:fedorov14,bib:petrovic13}, but also can substitute one of the carbon atoms in the hexagonal structure of graphene (usually nitrogen atom)\,\cite{bib:hage18}. In order to investigate the effects of electron doping on the high-energy optical absorption of graphene we introduce the Li atoms in one case and substitute carbon atoms with the N atoms in the other. The corresponding results of optical absorption are shown in Figs.\,\ref{fig:fig3}(a)-(c) and Figs.\,\ref{fig:fig3}(d)-(f), respectively, along with the temperature dependencies. The overall effect of the electron doping is the reduction of the intensity $I_{\pi}$ and the shift of the energy $\varepsilon_{\pi}$, which is in line with the increase of the EPC strength with doping. In fact, recent electron energy loss spectroscopy study showed that the doping via substituted N atom redshifts the VHs peak with respect to the pristine graphene\,\cite{bib:hage18}, in accordance with our results [e.g., see orange circles in Fig.\,\ref{fig:fig3}(d)]. Furthermore, the low-energy absorption is decreased below the two times Fermi energy due to the Pauli blocking. For the 1.54\% Li and 1.56\% N concetrations the additional absorption peak appears below 0.5\,eV that comes from the graphene-dopant interband transitions\,\cite{bib:novko17}. The temperature dependence is also slightly altered by presence of the dopant atom. For instance, the energy shift for 1.54\% Li-doped graphene is $\Delta_{\pi}\approx210$\,meV, while for the less-doped and pristine is $\Delta_{\pi}\approx150$\,meV. The differences for the other cases are present, but not so pronounced. We note that the energy of the VHs peak appears to be blueshifted for 1.56\% N-doped graphene [green circles in Fig.\,\ref{fig:fig3}(d)], but it is clear from Fig.\,\ref{fig:fig3}(f) that the shift of the overall absorption peak is negative with respect to the 0.78\% N-doped and pristine graphene.

\begin{figure}[t]
\includegraphics[width=0.48\textwidth]{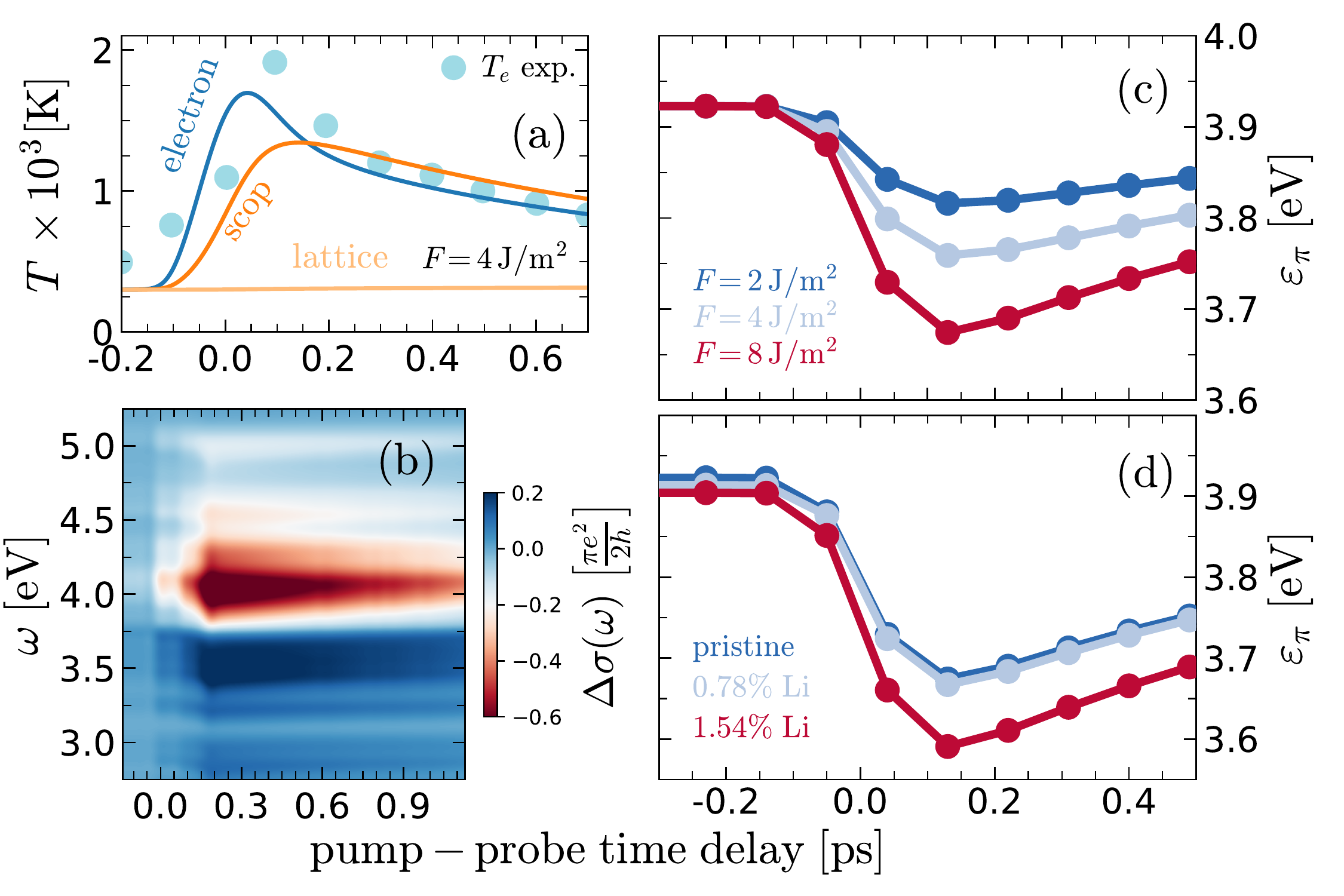}
\caption{\label{fig:fig4}Transient optical response of graphene. (a) Time evolution of the electron, strongly-coupled-optical-phonon, and lattice (representing the rest of the modes) temperatures when the fluence of the 100\,fs pump pulse is $F=4$\,J/m$^2$. Experimental values of electron temperature extracted from Ref.\,\cite{bib:johannsen13}, where the fluence of the 30\,fs pump pulse is 3.46\,J/m$^2$, are shown as well. (b) The corresponding differential absorption $\Delta\sigma(\omega)$ as a function of pump-probe time delay. Time-resolved shifts of the VHs peak $\varepsilon_{\pi}$ as a function (c) pump fluence $F$ and (d) electron doping concentration.}
\end{figure}

\textbf{Transient absorption.--} Now we use the optical absorption formula $\sigma(\omega)$ with phonon-assisted processes [Eq.\,\eqref{bib:eq1}] in combination with the effective temperature model\,\cite{bib:allen87,bib:lui10,bib:johannsen13,bib:caruso19} to reproduce the photoinduced optical response of graphene (see Methods for more details). The effective temperature model simulates the energy exchange between the pump laser pulse and the electrons in the material, and subsequent relaxation of electrons on relevant phonons via the electron-phonon scattering. Since the EPC in graphene is highly anisotropic (as we discussed above), we separate the total temperature into the three subsystems\,\cite{bib:lui10,bib:johannsen13,bib:caruso19}: the electrons, the strongly coupled optical phonons (the $E_{2g}$ and $A_1'$ modes), and the rest (lattice; mainly coming from acoustic phonons). In Fig.\,\ref{fig:fig4}(a) we show the obtained temperature change as a function of pump-probe time delay for these three subsystems when the pump fluence is $F=4$\,J/m$^2$ with 100\,fs duration time (i.e., typical experimental values\,\cite{bib:roberts14,bib:johannsen13}). The remnant phonon modes are so weakly coupled to the electronic system that the corresponding temperature is increased by around 30\,K in 1\,ps. Due to quite strong EPC the hot electrons transfer significant amount of their energy to the strongly coupled optical phonon modes already within first 100\,fs, whereby the temperature of these very few modes is elevated to about 1200\,K. Such a quick thermalization of hot electrons and hot strongly coupled optical phonons is experimentally well established\,\cite{bib:berciaud10,bib:kim15}. Hereafter electrons and strongly coupled optical phonons lose their energy to the reamining phonon modes via the weak electron-phonon and phonon-phonon couplings. This behaviour is in both qualitative and quantitative agreement with the results obtained with the time- and angle-resolved photoemission spectroscopy\,\cite{bib:johannsen13} (the small discrepancy between these two results for $0<t<100$\,fs mostly comes from the fact that we use laser duration of 100\,fs as in Ref.\,\cite{bib:roberts14}, while the experiment in Ref.\,\cite{bib:johannsen13} is done for 30\,fs). Note that we assume here that electrons are instantly thermalized and, thus, that the Fermi-Dirac distribution function with high electron temperature is well-defined quantity from the start. We base these assumptions on the results of the recent ultrafast time-resolved photoemission studies\,\cite{bib:rohde18,bib:tan17b}, where the electron thermalization time was determined to be less than 50\,fs.

Thus obtained temperatures are then utilized for simulating the transient optical absorption in graphene. Considering that the EPC-induced band structure renormalization in graphene arises predominantly from the $E_{2g}$ mode, while the rest of the modes give almost no contribution, we take the phonon temperature displayed in Fig.\,\ref{fig:fig4}(a) as the total phonon temperature entering the phonon-assisted optical absorption formula. The ensuing transient changes in the optical spectrum of graphene are shown in Fig.\,\ref{fig:fig4}(b) in the form of the time-resolved differential absorption. As already stated, the same spectral changes were observed in ultrafast pump-probe optical spectroscopy of graphene\,\cite{bib:roberts14} and graphite\,\cite{bib:pagliara11}. In fact, similar transients in the same energy range can be discerned from the results of the time-resolved electron energy loss spectroscopy study of graphite\,\cite{bib:carbone09}. However, the concomitant theoretical analysis based in first principles that can decipher the microscopic mechanisms underlying these changes was so far not provided. On the other hand, our present work offers the full microscopic picture. The photoinduced hot electrons transfer their energy mostly to the optical $E_{2g}$ phonon mode, which in turn redshifts the interband gap around the VHs point due to the strong EPC. The latter decreases and increases, respectively, the optical absorption around and slightly below the original energy of the VHs peak $\varepsilon_{\pi}$. In Ref.\,\cite{bib:roberts14} these changes were mostly attributed to the hot acoustic phonons, however, here we show that their coupling to the electrons is small and their temperature increase is minute, which leads to an insignificant impact on the band structure renormalization. Additionally,  Figs.\,\ref{fig:fig4}(c) and \ref{fig:fig4}(d) show how the band renormalization changes with increasing fluence $F$ and increasing electron doping concentration. Stronger $F$ produces larger phonon temperatures and thus bigger band shifts. Increase of the excess electrons amplifies the EPC\,\cite{bib:novko17} which as well results in larger band structure renormalization.

In principle, there is another, potential pathway to alter the EPC via the laser-induced hot electrons. Namely, ultrafast laser excitations modify the phonon dispersion of graphene, specifically the Kohn anomaly of the $E_{2g}$ mode\,\cite{bib:lazzeri06,bib:piscanec04}, which in turn could change the EPC (note that the EPC constant could be approximately expressed as $\lambda=N(\varepsilon_F)D/M\omega_{\rm ph}^2$, where $N(\varepsilon_F)$ is DOS at Fermi level, $D$ is deformation potential, $\omega_{\rm ph}$ is the relevant phonon frequency and $M$ is the corresponding mass). However, the photoinduced modifications of the $E_{2g}$ mode under conditions relevant for the present study ($\sim 10\,$cm$^{-1}$)\,\cite{bib:ishioka08,bib:yan09} are not enough to produce significant variations in the EPC.

\begin{figure}[b]
\includegraphics[width=0.48\textwidth]{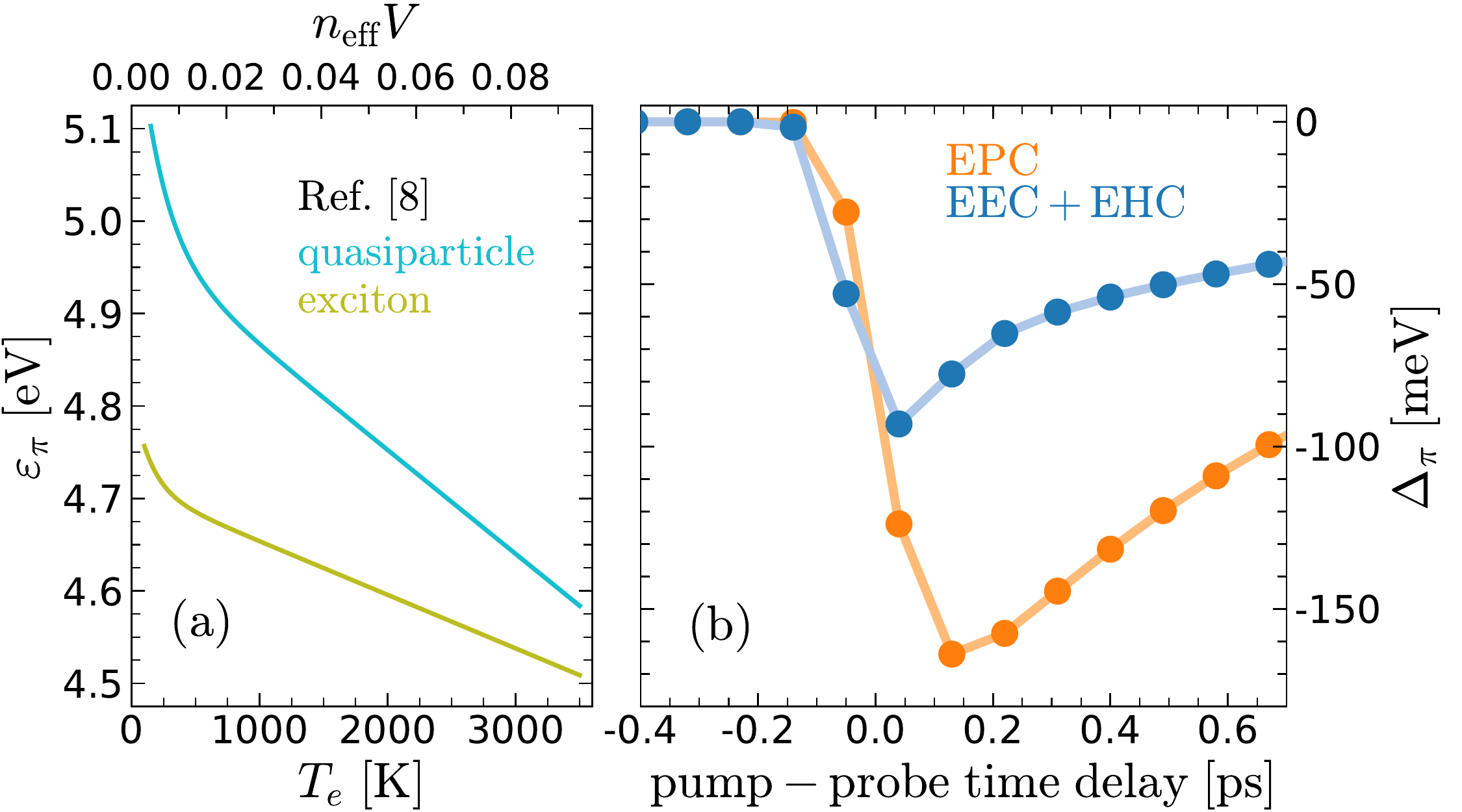}
\caption{\label{fig:fig5}Estimation of electron correlation effects on transient band structure renormalization. (a) Energy of the VHs peak $\varepsilon_{\pi}$ as a function of the excess electron concentration as obtained in Ref.\,\cite{bib:mak14} when the quasiparticle (turquoise) and additionally exictonic (green) effects are included. Dependence on the equivalent electron temperature $T_e$ is also shown. (b) Estimated energy shift of the VHs peak $\Delta_{\pi}$ due to the electron-electron and electron-hole couplings for $F=4$\,J/m$^2$ (blue). For comparison, the calculated $\Delta_{\pi}$ due to the EPC is shown in orange.}
\end{figure}

Finally, we discuss the impact of the Coulomb interactions, specifically quasiparticle (self-energy corrections) and excitonic (vertex corrections) effects, on the transient optical absorption in graphene. For equilibrium optical spectra of graphene it is well known that the aforesaid Coulomb interactions play significant roles\,\cite{bib:mak14,bib:yang11,bib:yang09,bib:kravets10,bib:mak11,bib:chae11,bib:malic11}. Besides, the rapid transient changes observed in ultrafast pump-probe optical spectroscopy of graphite (with decay time of 200\,fs) were ascribed to the electron-electron interactions, since the thermalization of the excited electron due to the latter interaction is generally much faster than due to EPC\,\cite{bib:pagliara11}. In Fig.\,\ref{fig:fig5}(a) we show the dependence of the VHs energy peak $\varepsilon_{\pi}$ on the excess electron concentration as obtained in Ref.\,\cite{bib:mak14} when the quasiparticle and additionally exictonic effects are included. When the electron temperature is elevated, the Fermi-Dirac distribution is skewed, and the number of thermally excited electrons and holes is increased. This allows us to match the dependence of $\varepsilon_{\pi}$ on the effective thermal charge density $n_{\mathrm{eff}}$ with the electron temperature $T_e$ (see Methods), and thus to the pump-probe time delay via the effective temperature model. The result of this qualitative estimation of the transient band structure renormalization due to the Coulomb interaction is depicted in Fig.\,\ref{fig:fig5}(b), along with the shifts due to the EPC. The estimation displays that both contributions can be of the same order of magnitude and in fact can have similar decay rates. All in all, the following scenario for the contribution of electron-electron interaction emerges. The laser enhances the concentration of thermally excited electrons and holes, increasing in turn the screening of electrons, which diminishes the quasiparticle and excitonic effects. The latter finally results in the redshift of the VHs peak (i.e., in the quenching of the excitonic effects in the VHs peak\,\cite{bib:chernikov15}). In the end, we remark that in order to provide a fully quantitative estimation of the impact of the Coulomb interactions on the photoinduced electron band shifts, one would need to go beyond the effective temperature model and simulate the nascent laser-induced electron distribution function (below 50\,fs)\,\cite{bib:rohde18} and the corresponding ultrafast quasiparticle and excitonic effects\,\cite{bib:pogna16,bib:perfetto15}. However, this goes beyond the scope of the present work.

\section*{Discussion}

In summary, by using the first-principles methodology we have studied the phonon-assisted processes in the photoinduced transient optical absorption of graphene. The results have shown that the inclusion of the electron-phonon coupling reduces the energy gap between occupied and unoccupied $\pi$ band around the Van-Hove-singularity point of the Brillouin zone. This is reflected in the optical absorption spectrum by the redshift and broadening of the corresponding Van-Hove-singularity peak. It is also shown that these phonon-assisted spectral changes are enhanced with elevated temperature, when the number of thermally excited phonons is increased, but also with the rise of the excess electron concentration, e.g., provided by the presence of the lithium and nitrogen atoms. To calculate the transient absorption spectra in graphene, we have simulated the laser-induced electron and phonon temperatures in good agreement with the time- and angle-resolved photoemission spectroscopy experiment\,\cite{bib:johannsen13}. The obtained transient differential absorption spectra have shown photoinduced reduction and increase of the absorption around and just below the original energy of the Van-Hove-singularity peak, which is again in a very good agreement with the recent pump-probe optical absorption experiments done on graphene\,\cite{bib:roberts14} and graphite\,\cite{bib:pagliara11}. These changes come from the band structure renormalization, and can be tuned with the increase of the power of the pump laser and with the electron doping techniques. Our qualitative analysis reveals that the phonon-assisted redshift of the Van-Hove-singularity peak in graphene is predominantly underlain by a single optical phonon mode concentrated at the center of the Brillouin zone, i.e., the $E_{2g}$ mode. Finally, we have made the qualitative estimation of the contribution coming from the electron-electron and electron-hole interactions and compared it to the electron-phonon coupling effects on the band shifts. The comparisons shows that the phonon-induced changes can even be larger than the ones triggered by the Coulomb interactions, which is at odds with the state-of-the-art beliefs. However, to reach a final, more quantitative, estimation on the latter, further investigations are needed.

We expect these results to be instrumental for gaining a full insight into electronic structure changes in graphene-based materials. For instance, for understanding the interaction between the image potential and $\pi$ states of graphite under nonequilibrium condition\,\cite{bib:pagliara13,bib:montagnese16,bib:tan17,bib:tan18}. Even more, the theoretical framework and the conclusions drawn here are not system-specific and can be helpful in elucidating photoinduced transient features in optical spectra of any material where phonon-assisted processes might also play important role, e.g., in emerging van der Waals heterostructures.


\section*{Methods}

\textbf{Phonon-assisted absorption.} To include the EPC effects into the optical conductivity Eq.\,\eqref{bib:eq1} we distort the carbon atoms $\kappa$ along the $\alpha$ direction in the supercell according to the following formula\,\cite{bib:zacharias16,bib:giustino17}
\begin{eqnarray}
\Delta\tau_{\kappa\alpha}=(M_{\kappa})^{-1/2}\sum_{\nu}(-1)^{\nu-1}e_{\kappa\alpha,\nu}\gamma_{\nu,T},
\end{eqnarray}
where sum goes over all phonon modes $\nu$, $e_{\kappa\alpha,\nu}$ are the eigenvectors obtained by diagonalizing the dynamical matrix, $M_{\kappa}$ is the mass of the atoms $\kappa$, and the magnitude of each displacement at temperature $T$ is $\gamma^2_{\nu,T}=(2n_{\nu}+1)\hbar/2\omega_{\nu}$. The Bose-Einstein distribution function is given with $n_{\nu}=1/(e^{\hbar\omega_{\nu}/k_b T}-1)$. The converged optical absorption spectra are obtained for the distorted $8\times 8$ supercells. We note that the final optical absorption spectra presented in the main text are an average of the optical absorption for parallel and perpendicular polarization directions.

\textbf{Effective temperature model.}
In this work we divide the total temperature into three subsystems\,\cite{bib:caruso19}, i.e., electron temperature $T_e$, temperature of the strongly coupled optical phonons (the $E_{2g}$ and $A_1'$ modes) $T_{op}$, and the temperature of the remaining lattice vibrations $T_l$. The interaction between these three subsystems is modeled with the following coupled rate equations\,\cite{bib:allen87,bib:johannsen13,bib:lui10,bib:caruso19}
\begin{eqnarray}
\frac{d T_{e}}{dt} &=& \frac{I(t)}{\beta C_{e}}- \frac{G_{op}}{C_{e}}(T_{e}-T_{op}) - \frac{G_{l}}{C_{e}}(T_{e}-T_{l}),
\\
\frac{d T_{op}}{dt} &=& \frac{G_{op}}{C_{op}}(T_{e}-T_{op}) - \frac{T_{op}-T_{l}}{\tau},
\\
\frac{d T_{l}}{dt} &=& \frac{G_{l}}{C_{l}}(T_{e}-T_{l}) + \frac{C_{op}}{C_{l}}\frac{T_{op}-T_{l}}{\tau},
\end{eqnarray}
where $G_{op}$ and $G_{l}$ are temperature-dependent electron scattering rates due to the strongly coupled optical phonons and the remaining lattice vibrations defined as,
\begin{eqnarray}
G_{\nu}=\frac{\pi k_B}{\hbar N(\varepsilon_F)}\lambda_{\nu}\left\langle \omega^2 \right\rangle_{\nu}\int_{-\infty}^{\infty} d\varepsilon N^2(\varepsilon)\left(- \frac{\partial f(\varepsilon;T_{e})}{\partial T_{e}} \right),
\end{eqnarray}
where $\lambda_{\nu}$ are the electron-phonon coupling strengths, $\left\langle \omega^2 \right\rangle_{\nu}$ are the corresponding second moment of the phonon spectrum, and $N(\varepsilon)$ is the electron density of states. These scattering rates are computed in the framework of density functional perturbation theory\,\cite{bib:baroni01,bib:lin08,bib:caruso19}.
$C_e$, $C_{op}$, and $C_{l}$ are the temperature-dependent specific heats of the aforesaid subsystems that are calculated by using electron and phonon density of states obtained with density functional and density functional perturbation theories as in Ref.\,\cite{bib:lin08,bib:caruso19}. $I(t)$ is the intensity of the laser pulse with Gaussian profile, characterized with the fluence $F$ and and the pulse duration. $\tau$ is the anharmonic scattering rate between the strongly coupled optical phonons and the rest of the modes, which is taken from the existing first-principles calculations\,\cite{bib:bonini07}. $\beta$ is the only fitting parameter of the model that controls the amount of the absorbed laser pump energy\,\cite{bib:johannsen13}. See Ref.\,\cite{bib:caruso19} for further details. 
For these calculations and thus for transient optical absoprtion calculations, we introduce small hole doping of about 0.0043 $h$ per unit cell (corresponding to $\varepsilon_F\approx-250$\,meV), because the estimated Fermi level in the experiment relevant for this study, i.e., Ref.\,[22], is around $-100$\,meV, and also because the experimental reference for our effective temperature model is the tr-ARPES study presented in Ref.\,[45], where the Fermi level is estimated to be $-240$\,meV. Since we don't expect significant quantitative and, especially, qualitative differences in our results between $\varepsilon_F=-250$\,meV and $\varepsilon_F=-100$\,meV, we use the former. Such graphene with a small hole doping is dubbed ``pristine'' in Fig\.\,\ref{fig:fig4}(d) in order to differentiate non-intercalated and Li-intercalated cases.
Note that one would need to go beyond the present effective temperature model in order to simulate the very earl stage of electron dynamics below 50\,fs and the accompanying nascent nonthermal electron distribution\,\cite{bib:sentef13,bib:waldecker16,bib:maldonado17,bib:rohde18}. However, this is beyond the scope of the present work, since we are interested here in moderate laser fluences and in the time scales larger than 50\,fs.

In order to simulate the transient optical absorption we combine the above effective temperature model and optical conductivity formula where phonon-asissted processes are included, i.e., Eq.\,\eqref{bib:eq1}. Similarly, the equilibrium optical conductivity formula, in which the total temperature is separated into effective electron and boson temperatures, was successfully used for describing nonequilibrium optical phenomena in cuprates in Refs.\,\cite{bib:dalconte12,bib:dalconte15}.

\textbf{Thermally excited charge.}
The effective number of conductive carriers, which usually enters the direct-current and optical intraband conductivity, is given with the following expression\,\cite{bib:novko16,bib:kupcic14}
\begin{eqnarray}
n_{\mathrm{eff}}\cdot V=\sum_{n\mathbf{k}}\left(-\frac{\partial f_{n\mathbf{k}}}{\partial \varepsilon_{n\mathbf{k}}}\right)\left| J_{n\mathbf{k}} \right|^2.
\end{eqnarray}
Since the Fermi-Dirac distrubtion function $f_{n\mathbf{k}}$ depends on the Fermi energy $\varepsilon_F$ and electron temperature $T_e$, the effective number of conductive carriers can be enhanced either by providing the external excess charge (e.g., via chemical doping) or by heating the system. The interband contribution to the effective number of conductive carriers is insignificant for small $\varepsilon_F$ and for $T_e=300-1600$\,K, as it is the case in the present work.

\textbf{Computational details.}
The ground state calculations are done by using the plane-wave based \textsc{quantum espresso} package\,\cite{bib:qe}, with a plane-wave cut-off energy of 60\,Ry. The core-electron interaction is approximated with norm-conserving pseudopotentials and the local density approximation (LDA) is used for the correlation functional. Phonon frequencies and eigenvectors for the distorted $8\times8$ supercells are calculated in the center of the Brillouin zone (the $\Gamma$ point) using density functional perturbation theory\,\cite{bib:baroni01}. The optical conductivity formula Eq.\,\eqref{bib:eq1} is calculated on a ($200\times200\times1$) and ($15\times15\times1$) Monkhorst-Pack grids for the undistorted $1\times 1$ and the distorted $8\times 8$ supercells, respectively. The numbers of the corresponding electronic bands included in the summation are 20 and 307, respectively. The broadening of the electronic states $\eta$ is 40\,meV for the $1\times 1$ and 80\,meV for the $8\times 8$ supercell. The current vertex functions $J^{\tau}_{nm\mathbf{k}}$ are calculated as in Ref.\,\cite{bib:novko16}. The electron-phonon scattering rates $G_{op}$ and $G_l$ are calculated by using the Eliashberg function, which is compute on $(300\times 300\times 1)$ and $(30\times 30\times 1)$ electron and phonon momentum grids, respectively. The associated Fermi energy is taken to be $\varepsilon_F=-200$\,meV as in the realistic graphene-substrate systems. The effective number of conductive carriers is calculated on a $(300\times 300\times 1)$ electron momentum grid and up to 4 unoccupied electronic bands are included.



\begin{acknowledgments}
Useful comments by Fabio Caruso are gratefully acknowledged. This work is supported from the European Regional Development Fund for the ``Center of Excellence for Advanced Materials and Sensing Devices'' (Grant No. KK.01.1.1.01.0001). D.N. also acknowledges financial support by Donostia International Physics Center (DIPC) during various stages of this work.
M.K. additionally acknowledges support from the Croatian Science Foundation (Grant No. IP-2016-06-3211).
Computational resources were provided by the DIPC computing center.
\end{acknowledgments}



\bibliography{piph}

\end{document}